\documentclass[twocolumn,trackchanges]{aastex631}
\usepackage{amsmath,booktabs,amssymb}
\usepackage{multirow}
\usepackage[version=4]{mhchem}
\usepackage{tabularx}
\usepackage{threeparttable}
\usepackage{booktabs}
\usepackage[version=4]{mhchem}
\usepackage{subfigure}
\usepackage{comment}
\begin{document}

% title -- completely open to suggestions
\title{Automated Assignment and Prediction of Molecules in Astronomical Line Surveys Using Machine-Learning-Based Chemical Embeddings} 

\author[0000-0001-5020-5774]{Zachary T. P. Fried}
\affiliation{Department of Chemistry, Massachusetts Institute of Technology, Cambridge, MA 02139, USA}

%%%%%%%%%%%%%%%%%%%%%%%%%%%%%%%%%%%%%%%%%
%%      Alphabetized GOTHAM LIST       %%
%%     Feel free to add yourself
%%%%%%%%%%%%%%%%%%%%%%%%%%%%%%%%%%%%%%%%%

\author[0000-0002-8932-1219]{Ryan A. Loomis}
\affiliation{National Radio Astronomy Observatory, Charlottesville, VA 22903, USA}

\author[0000-0001-9133-8047]{Jes K. J{\o}rgensen}
\affiliation{Centre for Star and Planet Formation, Niels Bohr Institute \& Natural History Museum of Denmark, University of Copenhagen,
\O{}ster Voldgade 5–7, 1350 Copenhagen K., Denmark}

\author[0000-0002-6667-7773]{Andrew Lipnicky}
\affiliation{National Radio Astronomy Observatory, Charlottesville, VA 22903, USA}

\author[0000-0003-2760-2119]{Ci Xue}
\affiliation{National Radio Astronomy Observatory, Charlottesville, VA 22903, USA}

\author[0000-0002-0332-2641]{Gabi Wenzel}
\affiliation{Department of Chemistry, Massachusetts Institute of Technology, Cambridge, MA 02139, USA}
\affiliation{Center for Astrophysics \textbar{} Harvard \& Smithsonian, Cambridge, MA 02138, USA}

\author[0000-0001-8134-5681]{Thomas H. Speak}
\affiliation{Department of Chemistry, University of British Columbia, Vancouver, BC, Canada}

%%%%%%%%%%%%%%%%%%%%%%%%%%%%%%%%%%%%%%%%%
%%        Senior Personnel LIST        %%
%%%%%%%%%%%%%%%%%%%%%%%%%%%%%%%%%%%%%%%%%
\author[0000-0001-9142-0008]{Michael C. McCarthy}
\affiliation{Center for Astrophysics \textbar{} Harvard \& Smithsonian, Cambridge, MA 02138, USA}

\author[0000-0003-1254-4817]{Brett A. McGuire}
\affiliation{Department of Chemistry, Massachusetts Institute of Technology, Cambridge, MA 02139, USA}
\affiliation{National Radio Astronomy Observatory, Charlottesville, VA 22903, USA}

%%%%%%%%%%%%%%%%%%%%%%%%%%%%%%%%%%%%%
% Corresponding author information
\correspondingauthor{Zachary T. P. Fried, Brett A. McGuire}
\email{zfried@mit.edu, brettmc@mit.edu}

\begin{abstract}
Modern radio telescopes generate vast amounts of observational data, offering valuable insights into the molecular composition of interstellar sources. Identifying the molecules within these datasets typically involves time-consuming and labor-intensive manual analysis. This paper presents an automated method for assigning molecules in interstellar line surveys. The algorithm operates in two main stages. First, it automatically determines key parameters of the data, including excitation temperature, line width, and source velocity. Next, it assigns the observed spectral peaks by evaluating the spectroscopic match of the molecular candidates along with analyzing their chemical relevance to the interstellar source. The chemical relevance is determined by leveraging machine-learning-based molecular embedding techniques to analyze the regions of chemical space occupied by the observed species. Following the line assignment, this information is then used to generate new molecular candidates that occupy the same regions of chemical space. These newly generated species serve as promising targets for further investigation in the observational data. The algorithm was validated on spectral line surveys of the dark molecular cloud TMC-1 and the star-forming region IRAS 16293-2422B. In both cases, it identified at least 67 molecular species, accounting for over 90 percent of the analyzed line intensity, in 17 minutes or less while maintaining a high level of accuracy.
\end{abstract}

\section{Introduction}

Understanding interstellar molecular inventories is crucial for astrochemical research. Molecules can provide unique insights into the chemical and physical conditions of interstellar sources, including  temperature \citep{walmsley_ammonia_1983,bergin_ch_1994}, density \citep{benson_survey_1989, van_der_tak_impact_1999, evans_physical_1999}, and ionization \citep{caselli_ionization_1998, caselli_deuterated_2002, mccall_enhanced_2003}. They can also be used to probe specific kinematic processes, such as outflows and shocks \citep{sch97,bac01}. The vast majority ($>$90\%) of interstellar molecular detections have been achieved using radio astronomical observations \citep{census_mcguire}. With advanced radio astronomy facilities such as the Green Bank Telescope (GBT), the Atacama Large Millimeter/submillimeter Array (ALMA), the Very Large Array (VLA), the IRAM 30m 
Telescope, and the Yebes 40m Radio Telescope, it is now possible to obtain extremely sensitive and high-resolution observations of interstellar sources across increasingly wide frequency ranges in a single observation, enabling simultaneous detection of numerous molecular transitions. Future upgrades, including ALMA’s planned Wideband Sensitivity Upgrade \citep{carpenter_alma2030_2023}, will further expand these broadband observational capabilities. As a result, it is not uncommon for spectral line surveys to contain thousands of distinct features.

While this wealth of observational data has resulted in a boom in newly detected interstellar species (which is increasing annually at a non-linear rate; \citealt{census_mcguire}), the assignment of these extremely dense line surveys can be a challenging task. The observed spectral features primarily correspond to rotational emission from gas-phase molecules and ions. The assignment of these surveys is traditionally  tackled by manually comparing the observed astronomical spectral peaks to rotational spectra cataloged in online databases such as the Cologne Database for Molecular Spectroscopy (CDMS;  \citealt{cdms_database}), the Jet Propulsion Laboratory Database (JPL; \citealt{jpl_database}), the Lille Spectroscopic Database (LSD; \citealt{motiyenko_lille_2025}), and Splatalogue \citep{splat}. However, due to the wealth of cataloged molecules, there are often a large number of molecular candidates in the various databases that have rotational transitions nearby (within experimental uncertainty) any observed peak in the observational data. Consequently, this assignment process is both labor intensive and time consuming. Therefore, the potential to automate this process is appealing, but is complicated by the need for an expert to often make informed decisions based on chemical intuition, rather than heuristics that are readily translatable to logic gates and boolean operators.

As our software capabilities continue to develop, the assignment and fitting of molecules in interstellar line surveys is constantly becoming more efficient and robust. Early workflows often relied on forward-modeling spectra under astrophysically relevant source conditions. Tools such as Weeds \citep{maret_weeds_2011} enabled users to simulate molecular signal, generally under LTE conditions, by computing line intensities and optical depths for prescribed physical parameters. These simulated spectra could then be overlaid on observational data to assess whether a molecule was present. In these workflows, however, source parameters such as column density, excitation temperature, line width, and source velocity were often determined through iterative manual adjustment rather than fully automated global fitting.

As contemporary broadband observational facilities began generating more dense, highly blended line surveys, the field shifted toward automated optimization packages and fitting frameworks such as GOBASIC \citep{rad_gobasic_2016}, XCLASS/MAGIX \citep{moller_extended_2017}, MADCUBA-SLIM \citep{martin_slim_2019}, SAMER \citep{el-abd_automated_2024}, and the MCMC functionality of molsim \citep{molsim}. Each of these frameworks integrates optimization routines that are able to determine best-fit source parameters and molecular column densities from the data. Despite these improved capabilities, these approaches still rely heavily on manual setup. For example, molecule selection generally remains manual, where the researcher must actively select the molecular catalogs or candidate species to include in the analysis. Thus, while these methods are still widely used and are powerful tools for determining best-fit parameters of specified molecules in interstellar line surveys, they are not primarily designed to automatically determine a complete molecular inventory without the user first defining the candidate species.

More recently, tools have begun to automate molecular line identification and spectral fitting. For example, SPECTUNER, introduced by \cite{qiu_spectuner_2025}, focuses on automated line identification by using XCLASS spectral modeling with a peak-matching loss function to identify and fit candidate molecules from spectroscopic databases. Spectuner-D1 \citep{qiu_spectuner-d1_2026} subsequently used deep reinforcement learning to accelerate automated spectral fitting by providing improved initial estimates for local optimization. Together, these studies represent important advances toward more automated spectral analysis.

In this paper, we add to this list of assignment tools and present a new automated method for assigning spectral line features and determining physical source parameters in radio astronomical observations without requiring an a priori molecular inventory. Beyond assignments alone, this approach builds on the work of \cite{fried_automated_2024} by evaluating the structural and chemical relevance of each candidate within the context of the known molecular inventory. This process enables the algorithm to capture aspects of the chemical reasoning often used in manual molecular assignments within a machine-based framework.

The method described here first automatically determines the Doppler velocity, molecular excitation temperature, and line width of the observational data and then assigns molecular carriers of the spectral lines. Subsequently, by investigating the regions of chemical space occupied by the assigned molecules, a rank-ordered list of new molecular candidates is generated based on their similarity to the detected inventory. The generated species represent strong candidates for targeted laboratory experiments that could facilitate their identification in observational data. This approach is evaluated using two spectral line surveys: the GBT Observations of TMC-1: Hunting Aromatic Molecules (GOTHAM) survey of the TMC-1 molecular cloud \citep{mcguire_early_2020,xue_tmc_2025} and the ALMA Protostellar Interferometric Line Survey (PILS) targeting the protostellar hot corino IRAS 16293-2422B \citep{jorgensen_alma_2016}.

\section{Factors Considered in Line Assignment}

The line assignment algorithm introduced in this paper was designed to closely emulate the manual assignment process. Therefore, it is first useful to discuss the factors that would be considered if a scientist were manually assigning the spectroscopic features using catalogs available in online databases. 

The first consideration is typically the spectroscopic match of the simulated rotational catalog to the observational data \citep{2005ApJ...619..914S, 2019ApJ...871..112X}. This involves simulating the rotational spectrum of the molecular candidate at the proper excitation temperature and investigating whether the frequencies and relative intensities of the rotational transitions match the observed peaks, after correcting for any Doppler shift caused by the line-of-sight velocity of an astronomical object ($v_{\rm lsr}$). It is also important to check if there are any strong simulated peaks that are missing in the observational data. 

However, especially when considering weak lines or line-confused observations, it is possible that the spectroscopic signal can be a somewhat feasible match for a molecule that is not present in the interstellar source, especially if that molecule has a small number of strong transitions in the observed frequency range. Thus, if there are several remaining molecular candidates for a certain line following the spectroscopic investigation, scientists often rely on their ``chemical intuition” to determine the most likely assignment. For astrochemical investigation, this reliance on ``chemical intuition” is reasonable, since the molecules detected in interstellar sources often occupy quite well-defined and homogeneous regions of chemical space. For example, the abundant detected species in warm star-forming regions such as IRAS\,16293-2422B are generally quite saturated and oxygenated molecules (i.e. \ce{CH3OH}, \ce{H2CO}, \ce{CH3OCH3}, \ce{CH3OCHO}, etc.) \citep{jorgensen_alma_2016,jorgensen_alma-pils_2018, drozdovskaya_ingredients_2019} that were likely formed on grain surfaces and subsequently sublimated into the gas phase upon the warming of the ices by the protostellar radiation. On the other hand, a cold prestellar source such as TMC-1 has a molecular inventory dominated by highly unsaturated species (i.e. \ce{C3N}, \ce{HC5N}, \ce{HC7N}, etc.) along with aromatic molecules \citep{lee_machine_2021, xue_tmc_2025}. Thus, if we were assigning a spectroscopic transition observed toward TMC-1, we would be inclined to assign the transition to an unsaturated species as opposed to an oxygen-bearing saturated molecule.

\section{Automated Line Assignment Process}

The following section outlines each step of the automated line assignment process. This method begins with determining the physical parameters of the astronomical object, followed by assigning the spectroscopic peaks.

\subsection{Source Parameter Determination}
\label{subsection:parameters}
In order to accurately assign the data, we must first determine the $v_{\rm lsr}$, excitation temperature ($T_{\rm ex}$), and line width ($\Delta V$). This is vital because if we assume an incorrect $v_{\rm lsr}$, the simulated spectra will be shifted in frequency, which would make the correct line assignment nearly impossible. Furthermore, employing an incorrect $T_{\rm ex}$ would alter the relative intensities of the simulated spectra, thereby increasing the risk of misassignment. While the algorithm incorporates methods to automatically determine these source parameters, they can also be manually provided by the user when such parameters are known in advance.

First, the line width (in both frequency and velocity) is determined by fitting a Gaussian profile to each of the lines in the spectrum and measuring the full-width half max (FWHM). The median of the determined line widths is used for the remainder of the analysis. A single line width is used for all molecules in the analysis. 

Next, we attempt to determine  the $v_{\rm lsr}$ and in the process also derive a best-fit $T_{\rm ex}$. To do this, we begin with a limited list of astronomically common molecules, such as \ce{CO}, \ce{NH3}, \ce{CH3OH}, \ce{N2H+}, \ce{HC3N}, etc. Then, for the strongest lines in the spectrum, the algorithm queries all sufficiently strong transitions from this list that are within approximately 250 km\,s$^{-1}$ of the observed frequency. For every candidate transition that passes some simple intensity checks, the $v_{\mathrm{lsr}}$ required to Doppler-shift the catalog transition to the observed frequency is computed and stored. As a note, while the default list of molecules used to determine the $v_{\mathrm{lsr}}$ is fixed, additional molecules that are known to be present in the data can also be inputted by the user for this step.  Additionally, we determined that a $v_{\rm lsr}$ search window of 250 km s$^{-1}$ is suitable for most Galactic sources; however, users may freely expand or restrict this interval in the code to suit their needs or optimize performance.

Following the investigation of every strong line, the algorithm searches for the most-likely $v_{\rm lsr}$ range using a sliding window approach. More specifically, a velocity window of width $\pm \Delta V$ is swept across the set of candidate velocities, and the number of stored values falling within each window is counted. Here, the sliding-window half-width is set equal to the linewidth determined from the Gaussian fit in the previous step ($\Delta V$). This window allows the peak-finding routine to tolerate small offsets in the measured peak frequency while restricting candidate matches to transitions that fall within approximately one observed linewidth. The densest velocity window (i.e., the interval containing the highest concentration of candidate velocities) is then selected. This method is expected to be robust if the spectrum contains a sufficiently large number of peaks, as there should be a molecular candidate with the correct $v_{\rm lsr}$ for most lines. In contrast, any other $v_{\rm lsr}$ values derived from incorrect transitions will occur by random chance. As a result, the window containing the correct $v_{\rm lsr}$ should be the most occupied.

Finally, after identifying the velocity window of width $\pm \Delta V$ that is most likely to contain the true $v_{\rm lsr}$, we refine this estimate to obtain a single best-fit value. We first select the molecules whose candidate transitions fall within the densest velocity cluster, and then apply nonlinear least-squares optimization (via \texttt{scipy}; \citealt{scipy}) of their simulated spectra (allowing the column density, excitation temperature, and $v_{\mathrm{lsr}}$ to vary) to determine the best-fit $v_{\mathrm{lsr}}$ along with the excitation temperature. The fitting routine is initialized with a user-provided excitation temperature guess, and the solution is constrained to vary by at most $\pm 100$ K from this initial value.

\subsection{Line Assignment}
Once the source parameters are determined, the line assignment process is initiated. First, the molsim package \citep{molsim} is used to determine the peaks in the spectrum along with the noise level of the observational data. For each peak, the CDMS, JPL, and LSD rotational spectroscopy databases are queried for the transitions that are at most 1/2 of the FWHM away from the center frequency. If the line width of the dataset is sufficiently narrow, catalogs from CDMS including hyperfine splitting are used. For a source with extremely narrow lines such as TMC-1, the frequency threshold is expanded past one FWHM since the cataloged rotational spectra can have uncertainties greater than the linewidth. For each of the determined molecular candidates, the rotational spectrum is simulated using molsim at the determined $T_{\rm ex}$, $\Delta V$, and $v_{lsr}$. The resulting peak frequencies and intensities are stored. 

Once all candidate molecules are determined through this database querying, the algorithm attempts to assign the lines. Each line can either be uniquely assigned, labeled as unidentified, or determined to have multiple potential carriers (possibly denoting a blended line). The three factors that are considered during the assignment process are the frequency match of the observed peak to the catalog transition, the relative intensity match between the simulated and observed spectrum, and the structural/chemical relevance of each molecular candidate.

\subsubsection{Spectroscopic Match}
For the frequency match, the score is determined by a simple linear scaling factor based on the difference between the velocity-shifted catalog frequency and the observed peak. We attempted scaling this scoring based on the frequency uncertainty in the spectroscopic catalog, but were unable to improve the rankings by doing this. Next, the relative intensity match is computed by first simulating the rotational spectrum of the molecular candidate to match the observed peak intensity. These spectral simulations are generated using molsim \citep{molsim} under the assumptions of LTE conditions, a single excitation temperature, uniform linewidth, and Gaussian line profiles. The algorithm then checks that no unrealistically strong transitions are predicted to be present in the spectrum (i.e. 10 times more intense than the strongest observed line) and that the other strong simulated lines of the molecule are present in the observational data around their expected intensity. The adopted threshold values are practically identical to those determined in the work of \citet{fried_automated_2024}. Fairly lenient intensity thresholds are required to account for potential uncertainties in optical depth, molecular excitation temperature, or line blending.

If the molecular candidate is isotopically substituted or vibrationally excited, the algorithm then performs some limited further analysis. For example, if the source is extremely cold, a molecule will be excluded if it is vibrationally excited, as there would likely be little to no population in that state. For most isotopically substituted species, the candidate score is down-weighted if the corresponding parent isotopologue has not already been assigned to a stronger transition. However, abundant species such as \ce{CO}, \ce{OCS}, and methanol are excluded from this down-weighting procedure because optical-depth effects can strongly alter relative line intensities, and isotopologues of these molecules can produce strong features in observational data. This treatment provides a simple first-order handling of isotopologues. Future work could improve this analysis by more explicitly comparing the predicted spectra of isotopically substituted species with those of their parent molecules.

\subsubsection{Structural Relevance Determination}

In order to mimic chemical intuition in an automated fashion, each molecule is also scored by its structural/chemical relevance based on the other molecules observed in the source. An earlier attempt to assess the likelihood of a molecule in a chemical mixture was presented in \citet{fried_automated_2024}. The method described in the present paper builds upon that work, with the updates detailed in the following paragraphs.

For this process, we rely on machine learning-based chemical embedding methods. These techniques create numerical vector representations of molecules that encode molecular information such as substructures, functional groups, and geometry \citep{wigh22}. By representing a molecule as a vector, it is mapped to a specific location in chemical vector space. Similar vector representations will be generated for molecules that are chemically or structurally comparable, and these molecules will therefore be nearby each other in this chemical vector space.  By identifying the regions of chemical space occupied by known mixture components, we can assess the chemical relevance of each new molecular candidate to the mixture.

Our previous approach to this problem relied on a graph-based architecture \citep{fried_automated_2024}. This graph consisted of approximately 300,000 molecules, with each molecule corresponding to a unique graph node. Molecules were then connected via bidirectional edges if their vector representations were sufficiently similar. The graph functioned as a ranking system where all initial weight was assigned to the known mixture components. Through an iterative process, weight was then transferred through the edge connections, with the transferred weight being diminished during each additional step. Therefore, molecules that were directly connected to the known mixture components (and therefore close in chemical vector space) received a greater amount of weight than the molecules separated by several edge connections. The resulting weight was utilized to assess the chemical relevance of each molecule in the graph to the mixture and was subsequently applied as an additional heuristic in the line assignment process.

Although this method performed well across various chemical mixtures, there remained opportunities for further improvement. These were mainly based on the non-continuous chemical space surface that was generated by the graph-based architecture. For instance, the distance threshold for an edge connection (i.e. how similar two molecules need to be in chemical vector space to be connected by an edge) placed certain constraints and limitations on the graph. To illustrate this, if the threshold value for an edge connection was set to 12 (meaning that two molecules were connected if their vector representations had a Euclidean distance less than 12), a molecule that was nearly identical to a known mixture component would have the same local connectivity as a molecule with a distance of 11.99. Moreover, a molecule that had a distance of 12.01 from a known mixture component would be ranked notably lower than the molecule with a distance of 11.99. These non-continuous edge constraints could lead to unwanted behavior in the rankings. Additionally, constructing a specific molecular graph restricted our analysis to the molecules within it. Consequently, for entirely new applications involving a significantly different set of molecules, the graph would need to be regenerated to contain a distinct list of species.

To improve this, we have re-worked our investigation of chemical vector space. Now, instead of a graph architecture, we have adopted a method similar to a weighted Gaussian-kernel density estimation and created a continuous chemical space surface with Gaussian weight centered on the vector representations of the known mixture components. More specifically, for every known chemical mixture component, a Gaussian peak is centered at that location in chemical vector space. The weight of each Gaussian is scaled by how isolated that molecule is within the detected set (i.e., higher individual Gaussian weights are assigned to more isolated peaks), allowing structurally isolated molecules to contribute more strongly rather than being completely overwhelmed by highly occupied regions of chemical space during the scoring. This prevents the analysis from overfitting to the densest regions of the detected inventory and helps preserve sensitivity to chemically distinct species. The resulting Gaussian surfaces for each known mixture component are then summed to generate a complete continuous surface representation of chemical space with weight distributed around the known mixture components. The weight of any point on the surface can then be calculated using the value of the probability distribution function derived from the combined Gaussian distributions at that point. Thus, the chemical/structural relevance of any molecular candidate in the mixture can be assessed by evaluating the weight of the surface at its corresponding point. 

The parameters of the Gaussian peaks (such as the width of each Gaussian) were determined via a hyperparameter search. The optimal values were determined in conjunction with several other hyperparameters, such as the global line assignment score, which will be discussed later. Our main approach for structural-relevance hyperparameter determination was to perform a line-by-line grid-search validation using the IRAS 16293B and TMC-1 datasets. In this validation, we omitted the frequency and intensity tests and evaluated only the structural-relevance scores. We tested various parameter combinations and identified those that most effectively up-weighted the correct molecular carriers while down-weighting incorrect candidates. These parameters were therefore expected to be the most likely to improve the reliability of transition assignments. The selected parameter values were then manually refined only slightly based on their performance across additional chemically distinct molecular species and smaller observational datasets. While it is a limitation of our analysis in this paper that these hyperparameters were determined primarily using the IRAS 16293B and TMC-1 datasets, these are among the most well-characterized observational studies and, importantly, are chemically distinct from one another. For example, the relative distances between the embeddings of cyanopolyynes and aromatic species, which are more prevalent in dark clouds, may differ substantially from those of hot-core molecules such as methanol and formaldehyde. Therefore, these datasets provided a reasonable basis for selecting parameters that should be applicable to additional hot-core and dark-cloud datasets.

\begin{figure}[h!] % Use figure* for a two-column figure
    \centering
    \includegraphics[width=\columnwidth]{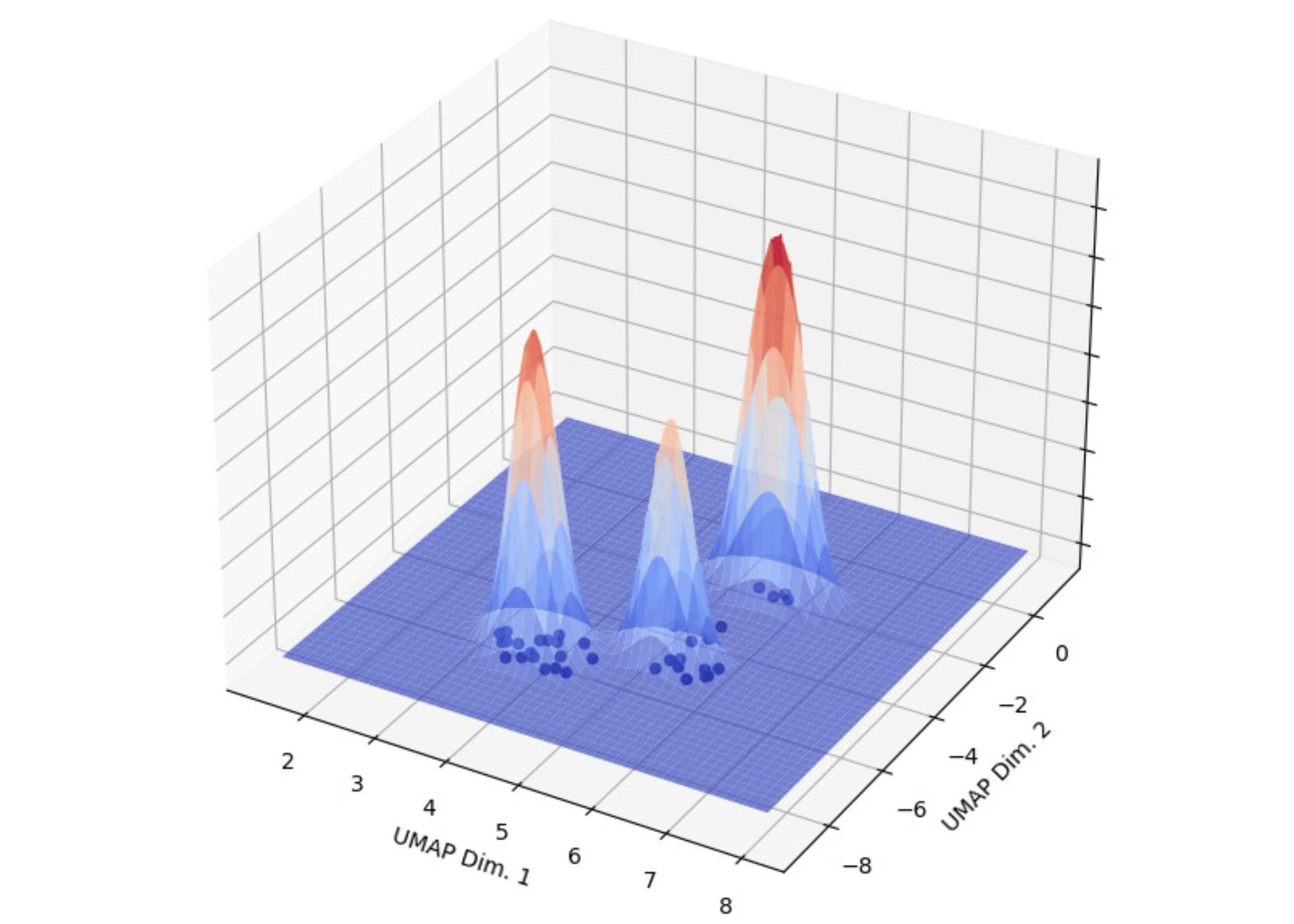} % Adjust width as needed
    \caption{Chemical space surface with Gaussian peaks centered on five abundant molecules in the IRAS 16293-2422B star forming region \citep{jorgensen_alma_2016}. The scatter points on the x-y axis are the UMAP 2-dimensional representations of all molecules detected in this source. It is important to note that because the embedding space is altered by the UMAP process, the width of the Gaussian in this figure does not accurately represent the width implemented by our code in 32-dimensional VICGAE space and is purely for illustration. The maximum intensity also varies across the plot because overlapping Gaussian peaks are summed.}
    \label{fig:surface}
\end{figure}

The surface no longer deals with the issues of non-continuity, since molecules that are closer to the peak of each Gaussian will have a higher ranking. Moreover, the architecture is not constrained to a pre-determined molecular graph since any point in chemical vector space can be sampled. Finally, while the graph-based ranking system required around 50 seconds to converge with optimal hyperparameters, this updated technique only takes a few seconds for the score of any molecule to be determined.

Figure~\ref{fig:surface} demonstrates a low-dimensional illustration of the updated structural-relevance metric and is intended for visualization purposes only. While the calculation in this work uses VICGAE molecular embeddings \citep{orion}, the general structure of the method should in principle be applicable to other molecular embedding methods that produce vector representations of different dimensionalities, although we have not explicitly tested other embedding methods in this work. In this figure, all molecules detected towards the star forming region IRAS 16293B are displayed as points in a scatter plot on the x-y axis \citep{jorgensen_alma_2016,jorgensen_alma-pils_2018,drozdovskaya_ingredients_2019}. A Gaussian peak has been centered on five abundant molecules in this source (namely \ce{OCS}, \ce{H2CO}, \ce{CH3OCH3}, \ce{CH3CHO}, and \ce{H2CCO}). As observed, most species in the scatter plot are already positioned in regions of high weight, even when conditioned on a small set of molecules.

At the beginning of the line assignment process, the user can choose to input a set of precursor molecules based on their prior knowledge of the source. For example, one could input methanol and formaldehyde for a star forming region. In this case, the chemical space surface is initialized with weight around these molecules. On the other hand, if they select not to input any precursor molecules, the first line is assigned solely using the scores from the spectroscopic match. The assigned molecule is then the first species which is given a peak on the chemical space surface.

The algorithm works in a sequential manner, in which the lines are assigned from strongest to weakest. After each new molecule is assigned, the weight surface is updated to incorporate this new information, and the score of any molecule can be recalculated. Since new information is gleaned with each new molecular assignment (and the chemical space surface becomes more informative), the previous assignments are all re-checked once a new molecule is identified. 

Furthermore, the algorithm accounts for the possibility of structurally unique molecules being present in any interstellar source by enabling an ``override" of the structural relevance scoring when sufficient spectroscopic evidence is available. Specifically, if a molecule has at least two lines in the data with a near-perfect spectroscopic match, and the structural relevance metric is the only factor reducing its molecular score, the molecule will be assigned to those lines. The species will then be added to the list of ``detected" molecules, and a peak will be marked at its position on the chemical space surface. This approach allows the algorithm to explore diverse regions of chemical space, avoiding constraints to a few narrow clusters.

For every molecular candidate of each line, a global score is tabulated by combining the frequency, intensity, and structural relevance scores. More specifically, the global score of each candidate is determined by first multiplying the structural relevance and frequency-offset scores using the following equation $S_{\rm global} = S_{\rm struct} \times S_{\rm freq}$, where $S_{\rm struct} \in [0,100]$ is the structural relevance percentile and $S_{\rm freq} \in [0,1]$ is the linear frequency-offset score. Next, multiplicative penalties of $\times 0.5$ are then applied for each failed intensity or isotopologue check. Therefore, a candidate for which there is a high-level of assignment confidence would have a score near 100. A softmax function is then applied to the global scores of all candidates for that line, generating a local score for each molecule relative to the other potential candidates. A molecule is confidently assigned to a line if its global and local scores exceed their respective optimized threshold values. As mentioned previously, these threshold values were determined in conjunction with various other hyperparameters. The optimized threshold values are extremely similar to those used in the automated line assignment described in \citet{fried_automated_2024}. If no molecule meets the global score threshold for a line, the transition is marked as unidentified. Conversely, if multiple molecules exceed the global score threshold but their scores are close enough that none surpass the local score threshold, the line is noted as having multiple possible carriers.

\subsubsection{Optimization of Column Densities and Spectral Model}
After every line is analyzed, we determine the column density of each molecule that best reproduces the observed spectrum, keeping the previously determined linewidth, $v_{\rm lsr}$, and excitation temperature fixed. This is accomplished using \texttt{scipy}'s nonlinear least-squares optimization \citep{scipy}, which adjusts the column densities of all assigned molecules simultaneously to produce the best-fit combined spectrum. During this process, molecules that were incorrectly assigned (typically those based on a single line whose full spectrum does not match the observations) are given very low column densities, resulting in a negligible contribution to the simulated spectrum. Molecules whose simulated spectra remain barely above the noise are then removed. This approach therefore allows the best-fit modeling to effectively identify and eliminate several false positive line assignments. Following the fitting, an interactive plot of the best-fit model as well as the determined column densities is provided to the user. 

\subsection{Molecular Prediction}
\label{sec:prediction}
Following the analysis of each line, we will have a chemical space surface that incorporates all of the assigned molecules in the source. This provides an informed understanding of the regions of chemical space occupied by the chemical inventory, enabling the prediction of other molecules that may be present. These molecular candidates can serve as starting points for identifying the molecular carriers of the unassigned transitions.

In our previous graph-based approach, once the line assignment was completed, we conditioned the graph on all assigned molecules and identified the highest-ranked unassigned molecular species within the graph. This unfortunately limited our consideration to molecules in the graph. Additionally, since the computational efficiency was inversely proportional to the size of the graph, we needed to select a graph size that contained a large enough number of molecular candidates without being prohibitively slow. With our new approach, we are no longer limited to the graph architecture and can therefore feasibly recommend almost any valid molecule.

To do this, since we no longer have a predetermined list of molecules to select from, we must develop a generative method to identify the most likely molecular structures. This requires creating an approach to extract specific structural information from dense numerical molecular vectors. To do this, we trained a multilayer perceptron (MLP) to predict molecular fragments from an inputted VICGAE vector. More specifically, we compiled a list of 326 molecular fragments, primarily consisting of those that make up the molecules in CDMS, supplemented with a few additional manually selected subgroups. We then created approximately 150,000 molecules by randomly combining any of these molecular fragments using the functionality of the Group SELFIES framework \citep{cheng_group_2023}. We then created VICGAE vector representations of each randomly generated molecule. For every molecule, we also produced a multi-hot encoded vector of length 326 that denotes which subgroups are present in the molecule. For example, if a benzene ring is contained in the compound, there is a value of one at the dimension corresponding to this fragment, otherwise the dimension has a value of zero.

With these two vectors for each molecule, we trained a MLP to map the 32 dimensional feature vector to the 326 dimensional multi-hot encoded vector. Thus, for any inputted feature vector, the model predicts which subgroups are present in the molecular species. The output layer of the model contains a sigmoid activation function, which results in every dimension of the output vector having a value between 0 and 1. This can therefore be interpreted as the probability that each fragment is present in a molecule given its VICGAE vector representation. The primary objective of this process is to identify the molecular constituents that are prevalent among molecules within a specific region of the chemical vector space. This trained MLP is more thoroughly described in Appendix~\ref{apendix_b}.

With this trained MLP, we can sample many of the highly weighted points on the chemical space surface following the line assignment and input the corresponding vectors into the network. To sample these points, we draw 90 vectors for each assigned molecule uniformly from a hypercube centered on that molecule's embedding vector. The width of each hypercube is set equal to the optimized Gaussian width. This focuses sampling around high-scoring regions of chemical space while still allowing local exploration around each assigned molecule. It also allows the model to sample regions that may receive high structural-relevance scores from the overlapping tails of multiple nearby Gaussians. The MLP will then predict the chemical makeup of a molecule in the sampled region of chemical space. For each point we randomly generate several molecules by sampling the molecular fragments in a probabilistic manner based on the MLP output and concatenating them into a full molecule with the Group SELFIES package \citep{cheng_group_2023}. At this point, we can also incorporate our domain knowledge of the source. For instance, the predicted molecules can be restricted to include only the atoms present in the assigned molecules. Additionally, we know that some classes of molecules, such as peroxides, are very rarely identified in radio astronomical observations, and can down-weight such species. In the current implementation, molecular filtering requires generated candidates to contain only atom types already present in the assigned molecular inventory. The algorithm also analyzes the assigned inventory to estimate the fractions of radicals, ions, and molecules containing non-terminal heteroatoms. Generated candidates with these properties are then retained probabilistically, with retention probabilities proportional to their occurrence in the assigned inventory.

\begin{figure*}[htb] % Use figure* for a two-column figure
    \centering
    \includegraphics[width=\textwidth]{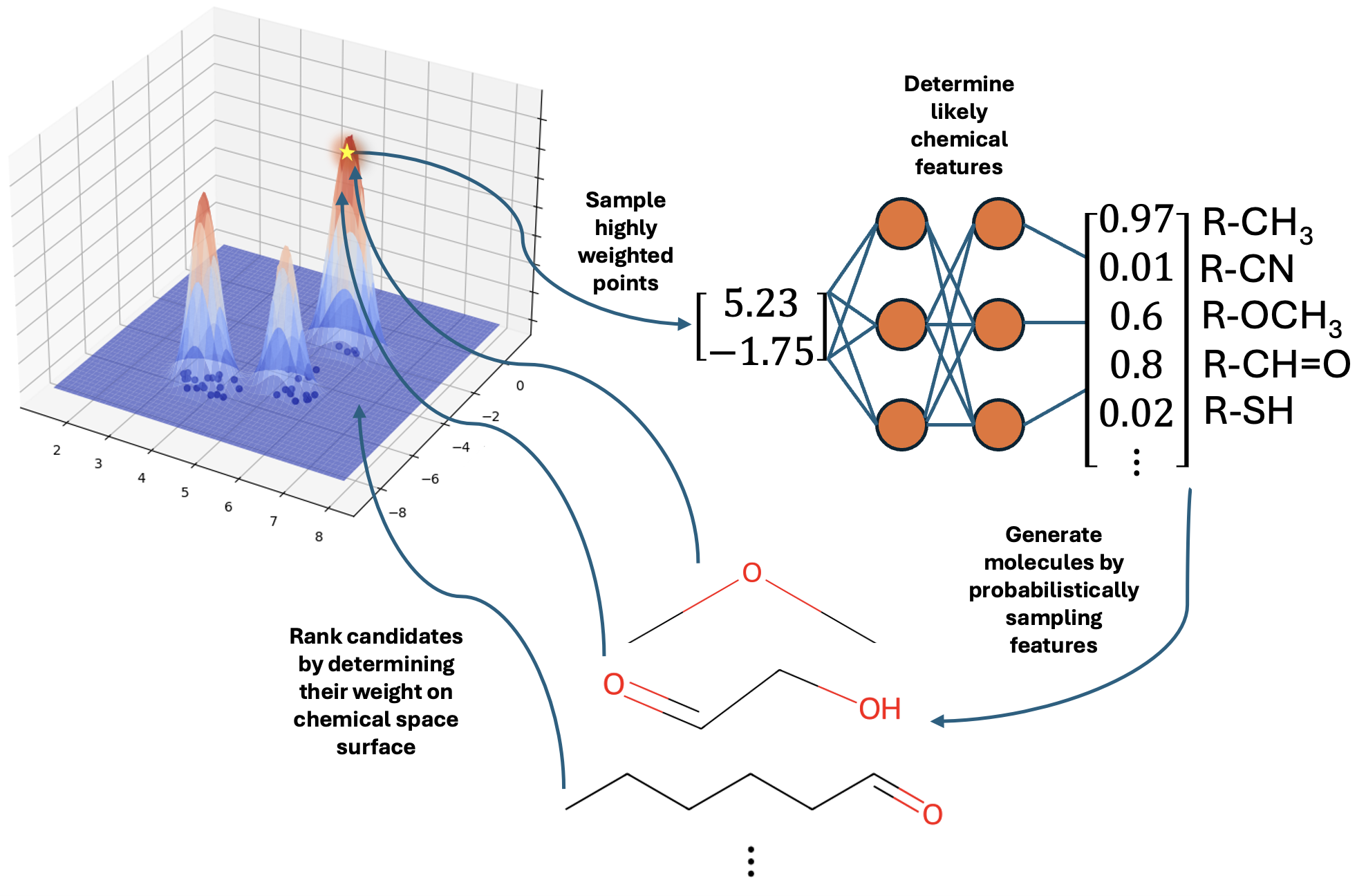} % Adjust width as needed
    \caption{Schematic of generative process to determine most chemically relevant molecular candidates for the unassigned spectroscopic transitions.}
    \label{fig:generalize}
\end{figure*}

We then produce VICGAE vector representations of this new set of molecular candidates, and finally determine which of these molecules have the greatest weight on the chemical space surface. The top ranked predicted molecules are therefore the most chemically relevant to the mixture and can be used as starting points in the identification of the unassigned transitions. This process is depicted in Figure~\ref{fig:generalize}. It is  important to note that while this method identifies strong molecular candidates for detection in interstellar sources based on chemical relevance, it represents an initial prediction step rather than a complete detection pipeline. Subsequent computational work and laboratory experiments are needed to accurately determine rotational constants, enabling detection of these molecules.

\section{Results}

\subsection{Line Assignment}
\subsubsection{TMC-1}
We have applied this algorithm on the spectral survey towards the cold molecular cloud TMC-1 from the fifth data reduction of the GOTHAM program \citep{mcguire_early_2020, xue_tmc_2025}. For this proof-of-concept investigation, we focused on the data ranging from 18 to 32 GHz. The algorithm analyzed 568 lines at $5\sigma$ significance. The peaks in the data were identified using the original GOTHAM dataset. However, the extremely high frequency resolution \citep[approximately 1.4 kHz, ][]{xue_tmc_2025} made it impractical to simultaneously fit the spectra of all assigned molecules in the original data. Consequently, for constructing the best-fit model, the spectrum was downsampled by a factor of eight.

The determined $v_{\rm lsr}$, linewidth, and excitation temperature are 5.8\,km\,s$^{-1}$, 0.31\,km\,s$^{-1}$, and 6.9\,K, respectively, which closely match the representative values for this source. This work assumes a single Gaussian component for each line feature, whereas the original GOTHAM observations resolve four components within a single line feature \citep{xue_tmc_2025}. Considering that each component in the four-component model has a line width of $\sim0.12\,{\rm km}\,{\rm s}^{-1}$ \citep[e.g.][]{2021NatAs...5..188L, 2021Sci...371.1265M}, our value of 0.31\,km\,s$^{-1}$ is consistent with the overall width of an observed line feature. The determined $v_{\rm lsr}$ is also consistent with the mean velocity of $\sim5.8{\rm km}\,{\rm s}^{-1}$ across the four components \citep{2021NatAs...5..188L}. The determined excitation temperature falls within the typical range of 5-10\,K reported for most molecules in this source \citep{2016ApJS..225...25G, xue_tmc_2025}.

%Moreover, Figure~\ref{fig:temp_sim} shows the rotational spectrum of H$^{13}$CCCCCN simulated at the determined parameter values overlaid on top of the GOTHAM observations. There is clearly a very high level of overlap between the simulated and observed spectrum, thus suggesting that the determined values are sufficiently accurate. 

Ultimately, including isotopologues, 67 molecules were assigned in the data. The collection of assigned molecules is listed in Table~\ref{tab:tmc_molecule_list} and displayed in Figure~\ref{fig:assigned_molecules}. To the best of our knowledge, all of these molecules have been previously detected towards TMC-1, thus suggesting that there were no ``false positive" assignments of molecules that are not truly present in the data. In the analyses of TMC-1 and IRAS 16293B, only catalogs of molecules composed of H, C, O, N, S, and Si were used. The code defaults to allowing H, C, O, N, and S, as these are by far the most common elements in detected interstellar species \citep{census_mcguire}; however, this selection can be easily modified by the user through an input parameter.

Out of the 568 spectral lines analyzed, 525 were successfully assigned to one or more molecular carriers, representing about 99\% of the total line strength of the identified $5\sigma$ lines, where the total line strength refers to the sum of peak intensities of all spectral lines above the $5\sigma$ threshold. The reported percentage is the sum of intensities of assigned lines divided by this total. This indicates that nearly all of the strong spectral features were assigned, and the fairly small number of remaining unassigned lines correspond to weak transitions.

\begin{table*}[htb!]
\centering
\begin{tabular}{c c c c c}
\hline
\ce{HC3N}, $(0,0,0,0)$ & \ce{HC5N}, $v = 0$ & \ce{CCS} & \ce{c-C3H2} & \ce{C4H}, $v = 0$ \\
\ce{SO}, $v = 0$ & \ce{HC7N}, $v = 0$ & \ce{NH3}, $v = 0$ & \ce{C3S}, $v = 0$ & \ce{CC\textsuperscript{13}CS} \\
\ce{C3N}, $v = 0$ & \ce{l-C4H2} & \ce{HCCNC} & \ce{H2CCN} & \ce{C6H}, $v = 0$ \\
\ce{DC3N}, $v = 0$ & \ce{HCC\textsuperscript{13}CN}, $v = 0$ & \ce{CH3CN}, $v = 0$ & \ce{HDCS} & \ce{CC\textsuperscript{34}S} \\
\ce{HC9N}, $v = 0$ & \ce{C2H3CN}, $v = 0$ & \ce{H\textsuperscript{13}CCCN}, $v = 0$ & \ce{HNC3} & \ce{C5H} \\
\ce{HC\textsuperscript{13}CCN}, $v = 0$ & \ce{HCCC\textsuperscript{15}N}, $v = 0$ & \ce{DC5N} & \ce{C3O}, $v = 0$ & \ce{l-C3H2} \\
\ce{HNCO} & \ce{CH3C4H} & \ce{CH3C3N} & \ce{HC3\textsuperscript{13}CCN} & \ce{HC2\textsuperscript{13}CC2N} \\
\ce{HC\textsuperscript{13}CC3N} & \ce{H\textsuperscript{13}CC4N} & \ce{HC4\textsuperscript{13}CN} & \ce{H2CCCHCN} & \ce{C3\textsuperscript{34}S} \\
\ce{SO2}, $v = 0$ & \ce{HCCCHO} & \ce{C6H-} & \ce{HC5NH+} & \ce{HC3O+} \\
\ce{H2CCO}, $v = 0$ & \ce{OCS}, $v = 0$ & \ce{c-H2C3O} & \ce{HCCS+} & \ce{HOCO+} \\
\ce{C8H} & \ce{HC5\textsuperscript{15}N} & \ce{C2O} & \ce{HSCN} & \ce{CH3C5N} \\
\ce{c-C6H5CN} & \ce{HC6\textsuperscript{13}CN} & \ce{HC3NH+} & \ce{HC2\textsuperscript{13}CC4N} & \ce{H\textsuperscript{13}CC6N} \\
\ce{HC5\textsuperscript{13}CCN} & \ce{HC3\textsuperscript{13}CC3N} & \ce{C5N} & \ce{l-C4HD} & \ce{C\textsuperscript{13}CCS} \\
\ce{HC\textsuperscript{13}CC5N} & \ce{DC7N} &  &  &  \\
\hline
\end{tabular}
\caption{List of molecules assigned in the GOTHAM data toward TMC-1. Note that the molecule formulas correspond to the entry titles of the CDMS, JPL, or LSD catalogs.}
\label{tab:tmc_molecule_list}
\end{table*}
%Furthermore, in three instances, a molecule that is not present in the data was only ruled out due to a low structural relevance score. For these molecules, the spectroscopic analysis did not provide sufficient evidence to dismiss them. This highlights the importance of incorporating structural relevance in the assignment algorithm, as the addition of these several misassignments would notably impair the algorithm's performance.

Of particular note, in six cases, molecules that were not actually present in the data could not be ruled out based solely on relative intensity analysis, yet were ultimately excluded in part due to their lower structural relevance scores compared to the assigned species. This correct exclusion of six potential misassignments suggests that the structural relevance analysis helped prevent potential false positive assignments, particularly for weak lines where assessing the quality of the spectroscopic match is more difficult. However, we observed that the algorithm occasionally struggled to initially assign high structural relevance scores to larger molecules within a chemical family. For example, \ce{HC9N} initially received a low structural relevance score despite the assignment of smaller cyanopolyynes such as \ce{HC3N}, \ce{HC5N}, and \ce{HC7N}. This indicates that, even with clear chemical similarity, notably increasing molecular size can in rare instances shift a molecule's position in the embedding space sufficiently to place it outside the regions occupied by most assigned species. While \ce{HC9N} was ultimately assigned correctly, this required the aforementioned ``override" capabilities due to substantial spectroscopic evidence. As the algorithm is applied to additional datasets, its hyperparameters can be refined to better capture such cases.

The algorithm was completed in 17.0 minutes on an Apple M2 Pro (10-core CPU, 16 GB RAM) using serial execution. Of this, 0.7 minutes were spent on source parameter ($v_{\rm lsr}$, $T_{ex}$, and $\Delta V$) determination, 0.5 minutes on catalog scraping and spectral simulation, 1.6 minutes on line assignment, and 14.1 minutes on the best-fit model of all assigned molecules. Of note, in this case the structural relevance metric was calculated a total of 21 times throughout the entire assignment process. The extended time required to obtain the best-fit model is a result of the exceptionally high resolution of the GOTHAM data. Even after downsampling by a factor of eight, the dataset still contains over 1 million datapoints. Consequently, the fitting process must evaluate the model and calculate residuals across all these points, making it a computationally intensive task.

\subsubsection{IRAS 16293B}
\label{subsec:iras_results}
To test the performance of this algorithm on data from a star-forming region, we utilized the ALMA PILS observations of IRAS 16293B in the frequency range of $\sim329 - 363$ GHz \citep{jorgensen_alma_2016}. These data are extremely line dense and have an extensive known molecular inventory \citep{jorgensen_alma_2016,drozdovskaya_ingredients_2019}. Moreover, this source is known to harbor enhanced isotopic substitution \citep{jorgensen_alma-pils_2018}, thus increasing the number of detectable isotopologues in the data. Although the PILS data exhibit narrow line widths by protostellar standards (around 1 km s$^{-1}$), the lines are still considerably broader than those in the GOTHAM observations, leading to more overlapping transitions. Additionally, the warmer temperatures allow for higher energy transitions and vibrationally excited states to be populated, which further increases the line-confusion and poses additional challenges for the algorithm.

For this dataset, our algorithm analyzed the 2,138 lines stronger than $5\sigma$ significance (with the noise level automatically determined by the code using the functionality of molsim \citep{molsim}). It is important to note that the algorithm calculates noise levels directly from the spectrum itself, which presents difficulties in highly line-dense data where identifying a clear noise floor becomes challenging. Consequently, the noise level automatically derived from the PILS spectrum by our code exceeds the RMS noise that can be measured from an off-source position in the ALMA data. Therefore, fewer spectral lines were analyzed in this study compared to those identified in earlier publications \citep{jorgensen_alma_2016}.  The line width was measured at 1.11 km s$^{-1}$, the $v_{\rm lsr}$ was found to be –0.16 km s$^{-1}$, and the excitation temperature was estimated to be 108 K. The data used for this analysis had already been velocity corrected, and therefore the determined $v_{\rm lsr}$ is in fairly close agreement with the expected value of 0 km s$^{-1}$. Additionally, the determined linewidth of 1.11 km s$^{-1}$ matches the value of 1 km s$^{-1}$ reported in \citet{jorgensen_alma-pils_2018}. Moreover, the molecules detected toward IRAS 16293B typically either have excitation temperatures in a ``cold" regime (100-150 K) or in a ``hot" regime (250-300 K) \citep{jorgensen_alma-pils_2018}. The fitted temperature in this case aligns with the lower range of the colder regime. This is somewhat surprising since most of the line-dense molecules used to determine the temperature and $v_{\rm lsr}$ in this source are those associated with hot excitation temperatures, such as \ce{CH3OH} and \ce{CH3OCHO} \citep{jorgensen_alma-pils_2018}. Therefore, in observational data of a protostellar source, optical depth effects are very likely influencing the relative intensities, which in turn affects the temperature determination. Also, the fit could likely be improved if the code were capable of incorporating multiple excitation temperatures. 

%Since the user-provided initial temperature guess was 150 K, this result corresponds to the upper limit of the allowed range, as the optimization is restricted to vary only within $\pm 100$ K of the input value. We have found that due to a combination of optical depth, overlapping lines, and nonstandard line shapes (i.e., self-absorption profiles), the determination of the excitation temperature using our least-squares fitting approach can be quite challenging. Thus, especially for these sources with wider line profiles, it is important to input an informative initial temperature guess. 

Ultimately, the algorithm assigned 80 molecules in the data, which are listed in Table~\ref{tab:iras_molecule_list} and displayed in Figure~\ref{fig:assigned_molecules_iras}. Nearly all of these appear to be correct assignments and have been previously reported toward IRAS 16293B, at least tentatively. However, one molecule, 
\ce{H2CN}, was assigned incorrectly. That being said, this species was only assigned to one weak blended line. Of note, this false-positive assignment disappears if we restrict the temperature to 250~K (corresponding to the hot regime in this source). Thus, future work will continue to focus on better handling optical depth such that the temperature determination is more robust and accurate. 

In total, 1,866 of the 2,138 analyzed transitions were assigned to one or more molecular carriers, representing 90.7\% of the total line strength of the identified $5\sigma$ lines. Excluding the signal corresponding to the strongest line---which as discussed later could not be properly assigned to \ce{CO} due to its self-absorption line profile---this fraction increases to $91.4\%$ of the $5\sigma$ intensity. As in TMC-1, the remaining unidentified lines are mainly weak features.

Moreover, there was one line where a molecule that was not the true carrier could not be excluded based on relative intensity analysis alone. However, its structural relevance score was sufficiently low to prevent incorrect assignment to an unidentified transition. This further suggests that structural relevance scoring generally improves assignment accuracy, particularly for weak lines that are difficult to assign based solely on spectroscopic analysis.

The primary isotopologues of \ce{CO}, HNC,  and HCN were incorrectly excluded from the assignments, although multiple isotopologues and/or vibrationally excited states were properly identified. This misassignment stems from the complex non-Gaussian line profiles in the strongest lines of these species due to optical depth, leading to significant discrepancies between the observed and simulated LTE spectra. This again highlights the limitations of assuming a Gaussian line profile for each species.

Ultimately, the assignment of the PILS data required a total of 5.8 minutes. Specifically, determining the $v_{\rm lsr}$, $T_{ex}$, and linewidth required 1.4 minutes, catalog scraping took 0.6 minutes, line assignment took 2.5 minutes, and generating the best-fit model took 1.3 minutes. Of note, in this case the structural relevance metric was calculated a total of 16 times throughout the entire assignment process.

\begin{table*}[htb!]
\centering
\begin{tabular}{c c c c c}
\hline
\ce{H2\textsuperscript{13}CO} & \ce{D2CO} & \ce{CH2DCHO}, $v_t = 0$ & \ce{H2CCO}, $v = 0$ & \ce{HDCO} \\
\ce{OCS}, $v = 0$ & \ce{H2C\textsuperscript{18}O} & \ce{CH3OCH3} & \ce{CH3CHO} & \ce{CS}, $v = 0 - 4$ \\
\ce{CH3OCHO}, $v_t=0,1$ & \ce{OC\textsuperscript{34}S} & \ce{\textsuperscript{13}CH3CHO}, $v_t \leq 1$ & \ce{CH3\textsuperscript{13}CHO}, $v_t \leq 1$ & \ce{c-C2H4O} \\
\ce{HC\textsuperscript{15}N}, $v = 0$ & \ce{HDS} & \ce{C\textsuperscript{18}O} & \ce{a-CH3OCH2D} & \ce{H\textsuperscript{13}CN}, $v = 0$ \\
\ce{C\textsuperscript{17}O} & \ce{SO}, $v = 0$ & \ce{H2C\textsuperscript{17}O} & \ce{CH3CN}, $v = 0$ & \ce{C\textsuperscript{34}S}, $v = 0, 1$ \\
\ce{CH3OD}, $v_t = 0$ & \ce{DCN}, $v = 0$ & \ce{CH3COCH3} & \ce{CH3OH}, $v_t = 0 - 2$ & \ce{HDCCO} \\
\ce{H2CS} & \ce{CH3CDO}, $v_t = 0, 1$ & \ce{s-CH3OCH2D} & \ce{DC(O)OCH3} & \ce{O\textsuperscript{13}CS} \\
\ce{CH2DOH} & \ce{CH2DCN} & \ce{CHD2OH}, $v_t = 0$ & \ce{CH3SH}, $v_t \leq 2$ & \ce{CH3CCH}, $v = 0$ \\
\ce{CH3\textsuperscript{13}CN}, $v = 0$ & \ce{CD3OH}, $v_t = 0$ & \ce{\textsuperscript{13}CH3OH}, $v_t = 0, 1$ & \ce{HNCO} & \ce{C2H5CN}, $v = 0$ \\
\ce{\textsuperscript{13}CH3CN}, $v = 0$ & \ce{HC(O)OCH2D} & \ce{C2H5OH} & \ce{CH3CN}, $v_8 = 1$ & \ce{OC\textsuperscript{33}S} \\
\ce{SiO} & \ce{CH3OD}, $v_t = 1$ & \ce{CH3NC} & \ce{HCOCH2OH} & \ce{SO2}, $v = 0$ \\
\ce{D2\textsuperscript{13}CO} & \ce{HC(O)OCHD2} & \ce{C\textsuperscript{33}S}, $v = 0, 1$ & \ce{c-C3H2} & \ce{\textsuperscript{13}CH3OCH3} \\
\ce{H2C\textsuperscript{13}CO} & \ce{c-C2H3DO} & \ce{HCN}, $v_2 = 1$ & \ce{CH3\textsuperscript{18}OH}, $v_t \leq 2$ & \ce{a-a-CH2DCH2OH} \\
\ce{HC3N}, $(0,0,0,0)$ & \ce{H\textsuperscript{13}C\textsuperscript{15}N}, $v = 0$ & \ce{CH3C\textsuperscript{15}N}, $v = 0$ & \ce{H2\textsuperscript{13}CCO} & \ce{CHD2CHO}, $v_t = 0$ \\
\ce{s-C2H5CHO}, $v = 0$ & aGg'-\ce{(CH2OH)2} & \ce{HN\textsuperscript{13}CO} & \ce{D\textsuperscript{13}CN} & \ce{NO}, $v = 0$ \\
\ce{H2CN} & \ce{HDC\textsuperscript{18}O} & \ce{a-CH3CHDOH} & \ce{2,2-c-CD2CH2O} & \ce{CH2(OH)CHO}, $v_{12} = 1$ \\
\hline
\end{tabular}
\caption{List of molecules assigned in the PILS data toward IRAS 16293-2422B. Note that the molecule formulas correspond to the entry titles of the CDMS, JPL, or LSD catalogs.}
\label{tab:iras_molecule_list}
\end{table*}

\subsubsection{Challenges In Line Assignment}
The datasets tested in this proof-of-concept work are both observational studies with significant broadband coverage and fairly narrow linewidths ($\lesssim 1$ km\,s$^{-1}$). Together, these characteristics enhance the accuracy of the line assignments, as narrow linewidths mitigate line confusion and the extensive frequency coverage ensures that each candidate molecule generally contains multiple transitions within the observed range, supporting relative intensity analysis. However, the algorithm is more prone to false-positive assignments in data that has limited frequency coverage or is extremely line confused. In such data, it is much more difficult to rule out molecular candidates, as missing transitions are harder to identify when most channels contain molecular lines and only a few transitions of each species are covered. Therefore, in these cases, it is likely that greater manual verification of the results is required. However, if there are persistent false-positive assignments in the data, the user is able to input a list of molecules that the code will be forced to ignore. 

Moreover, the present analysis includes only transitions with strengths above the $5\sigma$ threshold. Assigning lines weaker than this is not recommended, as automated evaluation of relative line intensities becomes unreliable for faint transitions, increasing the likelihood of false-positive assignments.

Additionally, the algorithm currently assumes a single $v_{\rm lsr}$, excitation temperature, and linewidth for all molecules in the survey, as well as Gaussian line profiles. In practice, these assumptions are often violated in real observational data. Future work may therefore focus on increasing the flexibility of the source parameters implemented in the code.

\subsection{Molecular Prediction Results}
\subsubsection{TMC-1}
When conditioned on the list of assigned species in the TMC-1 dataset, the predictive method produced 18,702 molecules. However, after removing duplicates and invalid molecules based on the provided chemical criteria (with the criteria for invalid species being outlined in Section~\ref{sec:prediction}), and the assigned species, 2,411 species remained. The top 100 ranked molecular candidates are displayed in Figure~\ref{fig:predicted_molecules_tmc}. This predictive process took only a few seconds following the line assignment. As can be seen, there are a relatively large number of duplicated predicted molecules because the fragment assembly is performed by token sampling based on the probabilities output by the trained MLP. Since molecules prevalent in interstellar sources generally contain a great deal of common substructure, the MLP often predicts higher probabilities for these common features, leading to these fragments being sampled more frequently. Thus, because we are commonly sampling and concatenating these highly predicted tokens, this results in a great deal of duplication in the predictions. By visually comparing Figure~\ref{fig:assigned_molecules} and Figure~\ref{fig:predicted_molecules_tmc}, it is evident that the algorithm is generating predominately unsaturated carbon-chain molecules that are quite similar to the detected species. A quantitative analysis of the chemical characteristics of the detected and top 100 ranked predicted molecules is displayed in Table~\ref{tab:molecular_predictions}. As can be seen, the top-ranked predicted molecules are on average almost identical in size to the observed molecular species. They are also somewhat similar in saturation, as both datasets are dominated by alkyne-containing species. Similar to the observed species, the predicted molecules also have a greater proportion of nitrogen than oxygen and sulfur.

Of note, the top ranked generated molecule, \ce{CH2CHCCH}, has been detected in TMC-1 with observations at frequencies higher than those of the GOTHAM observations \citep[35-47 GHz, ][]{cernicharo_ch2chcch_2021}. Several other molecules among the top-100 predictions in Figure \ref{fig:predicted_molecules_tmc} have also been detected in TMC-1, including some present in the GOTHAM data, though their strongest transitions lie outside the frequency range examined here. These include \ce{CH3CCH} \citep{irvine_new_1988}, \ce{HCCCHS} \citep{cernicharo_sulphur_2021}, \ce{HCCCH2CCH} \citep{fuentetaja_discovery_2024}, \ce{HCCO} \citep{cernicharo_discovery_hcco_2021}, \ce{CH3CH2CCH} \citep{cernicharo_ch3ch2cch_2024}, \ce{CH2(CN)2} \citep{agundez_rich_2024}, \ce{H2CCS} \citep{cernicharo_sulphur_2021}, \ce{HCCCH2CN} \citep{mcguire_early_2020}, \ce{H2CO} \citep{henkel_h2co_1981}, \ce{CH2CHCH3} \citep{marcelino_propylene_2007}, \ce{HOCN} \citep{cernicharo_hc3o_2020}, \ce{CH3SH} \citep{agundez_thioacetaldehyde_2025}, and \ce{CH3OH} \citep{soma_methanol_2015}. Moreover, several nonpolar molecules which are clearly relevant to TMC-1 were also highly-predicted, such as \ce{HC4H}, \ce{NC4N}, \ce{HCCH}, \ce{NCCN}, and \ce{C2H4}.

%\ce{CH2CHC3N} \cite{kelvin_lee_discovery_2021_cyanovinyl},\ce{HSCN} \citep{cernicharo_hocn_2024}, , and \ce{HOCN} \citep{cernicharo_hocn_2024}

These results indicate that the generative method effectively produces molecular candidates that are chemically relevant to the interstellar source. Although the algorithm is capable of feasibly generating virtually any valid molecule, out of the 18,702 molecules it produced, 3,232 corresponded to molecules assigned in TMC-1, including repeated instances. This outcome demonstrates that the model is successfully learning the functional groups and structural motifs that characterize molecules occupying specific regions of chemical space, thereby generating species that are particularly relevant to the interstellar environment.

\begin{comment}
This all indicates that the generative method can effectively produce molecular candidates pertinent to the interstellar source. Out of the 19,164 molecules generated by the algorithm (including duplicates), 3,259 matched molecules identified in TMC-1 during the line assignment process. This is a remarkably large percentage, as the method is capable of feasibly generating any valid molecule. The high percentage of predicted species that correspond to the detected molecules suggests that the MLP effectively predicts the correct functional groups for the sampled regions of chemical space. Furthermore, this once again indicates that the method is generating molecules that are highly relevant to the interstellar source. 
\end{comment}
\subsubsection{IRAS 16293B}

When conditioned on the molecules assigned toward IRAS 16293B, after filtering duplicates, assigned molecules, and invalid species (with the criteria for invalid species being outlined in Section~\ref{sec:prediction}), the code produced 1,464 unique molecular candidates (12,933 total without filtering). The top 100 ranked species are shown in Figure~\ref{fig:predicted_molecules_iras}. Among the 12,933 generated molecules, 4,619 matched those previously assigned by the algorithm (including duplicates), further indicating that the trained MLP effectively learns the functional groups characteristic of the source. The chemical characteristics of the assigned and predicted molecules toward IRAS 16293B are displayed in Table~\ref{tab:molecular_predictions}. As seen here, similar to the detected species, the molecules predicted toward IRAS 16293B are notably more saturated than those toward TMC-1. Moreover, the predicted molecules toward IRAS 16293B are also more oxygen-rich, consistent with the observational results. However, the predicted molecules toward this source appear to be more nitrogen-bearing than those actually assigned.

Additionally, several of the top-100 ranked predicted molecules have been detected in the PILS data corresponding to lines weaker than those investigated in this work (due to the fairly high noise level determined by our code, as discussed in Section~\ref{subsec:iras_results}). Examples include \ce{C3H6} \citep{manigand_alma-pils_2021} and \ce{CH2NH} \citep{ligterink_amines_2018}. 

\begin{table*}[ht]
\centering
\caption{Comparison of molecular properties between assigned and top-100 predicted molecules in both sources. The number of heavy atoms refers to non-hydrogen atoms. The degree of unsaturation quantifies the number of rings and multiple bonds in a molecule, calculated as (2C + 2 + N - H - X)/2, where C, N, H, and X represent the number of carbon, nitrogen, hydrogen, and halogen atoms, respectively. This denotes how many degrees a molecule deviates from being fully saturated (having only single bonds).}
\label{tab:molecular_predictions}
\begin{tabular}{lcc|cc}
\hline
\textbf{Property} & \multicolumn{2}{c|}{\textbf{TMC-1}} & \multicolumn{2}{c}{\textbf{IRAS 16293B}} \\
\cline{2-5}
 & \textbf{Assigned} & \textbf{Predicted} & \textbf{Assigned} & \textbf{Predicted} \\
\hline
Mean Degree of Unsaturation & 3.89 & 3.17 & 1.32 & 1.56 \\
Mean Number of Oxygens & 0.30 & 0.26 & 0.74 & 0.53 \\
Mean Number of Sulfurs & 0.19 & 0.31 & 0.23 & 0.41 \\
Mean Number of Nitrogens & 0.47 & 0.42 & 0.26 & 0.60 \\
Mean Heavy Atoms & 4.35 & 4.72 & 2.81 & 4.11 \\
\hline
\end{tabular}
\end{table*}

\section{Generalization of Structural Relevance Analysis}

In this paper, we analyzed the performance of the structural relevance metric and molecule generation technique on the chemical inventory of an interstellar source. That being said, the approach is broadly applicable across various domains. The primary aim of this methodology is to identify regions of chemical space occupied by molecules with specific properties. By targeting these regions, we can automatically generate new molecules that share those properties.

For astrochemists, the properties of interest are the abundance and detectability of molecules within a given interstellar region. However, the same approach could be extended to other fields, such as pharmaceutical science, by conditioning the chemical space surface on molecules with desired chemical or pharmaceutical properties. This is broadly analogous to property-guided molecular generation in pharmaceutical cheminformatics, where molecular generation can be biased toward desired objectives such as structural similarity or predicted biological activity, although the objective and implementation here are distinct \citep{olivecrona_molecular_2017}. Without any additional knowledge, the method can generate previously unexplored molecules nearby the input species in chemical vector space, making them promising candidates for exhibiting similar properties. These predicted molecules could then be further investigated through calculations or experiments to assess their relevance for a given application. Moreover, this technique could be combined with complementary experimental methods such as mass spectrometry, NMR, or FT-IR for applications including environmental water and air quality monitoring.

Additionally, we utilized the VICGAE embedding method in our work due to its success on prior astrochemical studies \citep{orion, toru_shay_exploring_2025}. However, the framework is flexible and can accommodate other embedding models optimized for different applications. Adapting the method would only require retraining the MLP and fine-tuning the parameters of the chemical space surface, such as the widths of the Gaussian peaks.

\section{Conclusion}

This paper presents a method for automatically assigning molecular species in radioastronomical observations. The approach begins by determining key source parameters, including excitation temperature, linewidth, and $v_{\rm lsr}$. It then assigns spectral lines by querying spectroscopic databases for molecules with known rotational transitions near the observed peak frequencies. For each molecular candidate, spectral simulations are performed using the molsim package to evaluate the match between the rotational catalog and the observational data. Additionally, the chemical and structural relevance of each molecule to the interstellar source is assessed by examining its position within the regions of chemical space occupied by the other assigned species. After analyzing each of the spectral lines, new molecular candidates are generated from the highly weighted regions of chemical space, offering potential starting points for identifying the carriers of the remaining unassigned transitions.

We demonstrated the performance of the algorithm by testing it on the GOTHAM observations of TMC-1 as well as the PILS observations of IRAS 16293B. The determined excitation temperature, linewidth, and $v_{\rm lsr}$ closely aligned with accepted literature values for both sources. The algorithm assigned 67 distinct species toward TMC-1 and 80 toward IRAS 16293B with only one false positive assignment. The entire assignment process was completed in around 17 minutes for the GOTHAM data and 6 minutes for the PILS survey. Furthermore, the top-ranked generated molecules closely matched the chemical characteristics of both detected inventories, even including some additional known molecules in the sources.

%Future work will combine this assignment process with a fitting method such as XCLASS \citep{moller_extended_2017} or SAMER \citep{el-abd_automated_2024} to determine the column densities of the assigned molecules. We also hope to develop additional functionality to identify several velocity or temperature components within the observational data.

\section{Data access \& code}
The code is publicly available with installation and usage instructions on \dataset[GitHub]{https://github.com/zfried/astro_amase} and \dataset[Zenodo]{https://doi.org/10.5281/zenodo.21169289}. The database and model files required to run the code, including the CDMS/JPL catalogs, precomputed molsim objects, and trained neural network weights, are available on \dataset[Zenodo]{https://doi.org/10.5281/zenodo.21171085}. The training datasets are available in a separate \dataset[Zenodo]{https://doi.org/10.5281/zenodo.21171237} repository.


\begin{acknowledgements}

\end{acknowledgements}

\software{
    Molsim \citep{molsim},
    Numpy \citep{numpy},
    PyTorch \citep{pytorch},
    Rdkit \citep{rdkit},
    Scipy \citep{scipy},
    Astropy \citep{astropy},
    Group Selfies \citep{cheng_group_2023},
    VICGAE \citep{orion}.
          }

\bibliography{references}

\appendix

\renewcommand\thefigure{\thesection\arabic{figure}}   
\renewcommand\thetable{\thesection\arabic{table}}    

\setcounter{figure}{0}    
\setcounter{table}{0} 

\section{Appendix A: Assigned and Predicted Molecules}
\label{apendix_a}

\begin{figure*}[htb] % Use figure* for a two-column figure
    \centering
    \includegraphics[width=\textwidth]{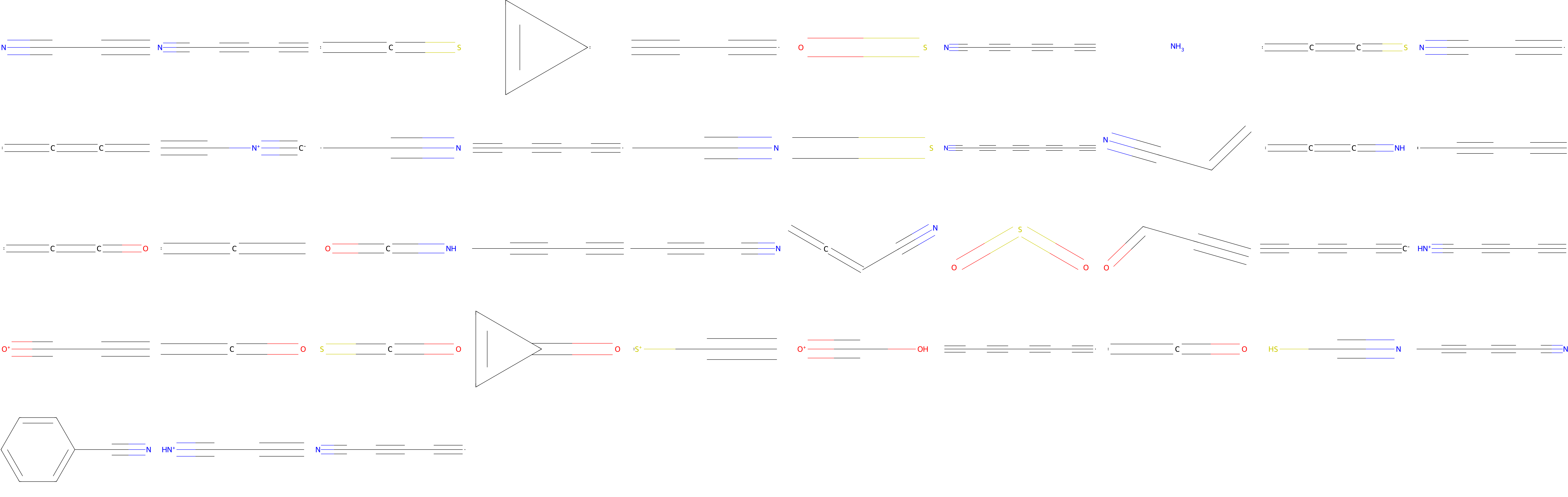} % Adjust width as needed
    \caption{All molecules assigned by the algorithm in the GOTHAM observations of TMC-1, not including isotopically substituted species.}
    \label{fig:assigned_molecules}
\end{figure*}

\begin{figure*}[htb] % Use figure* for a two-column figure
    \centering
    \includegraphics[width=\textwidth]{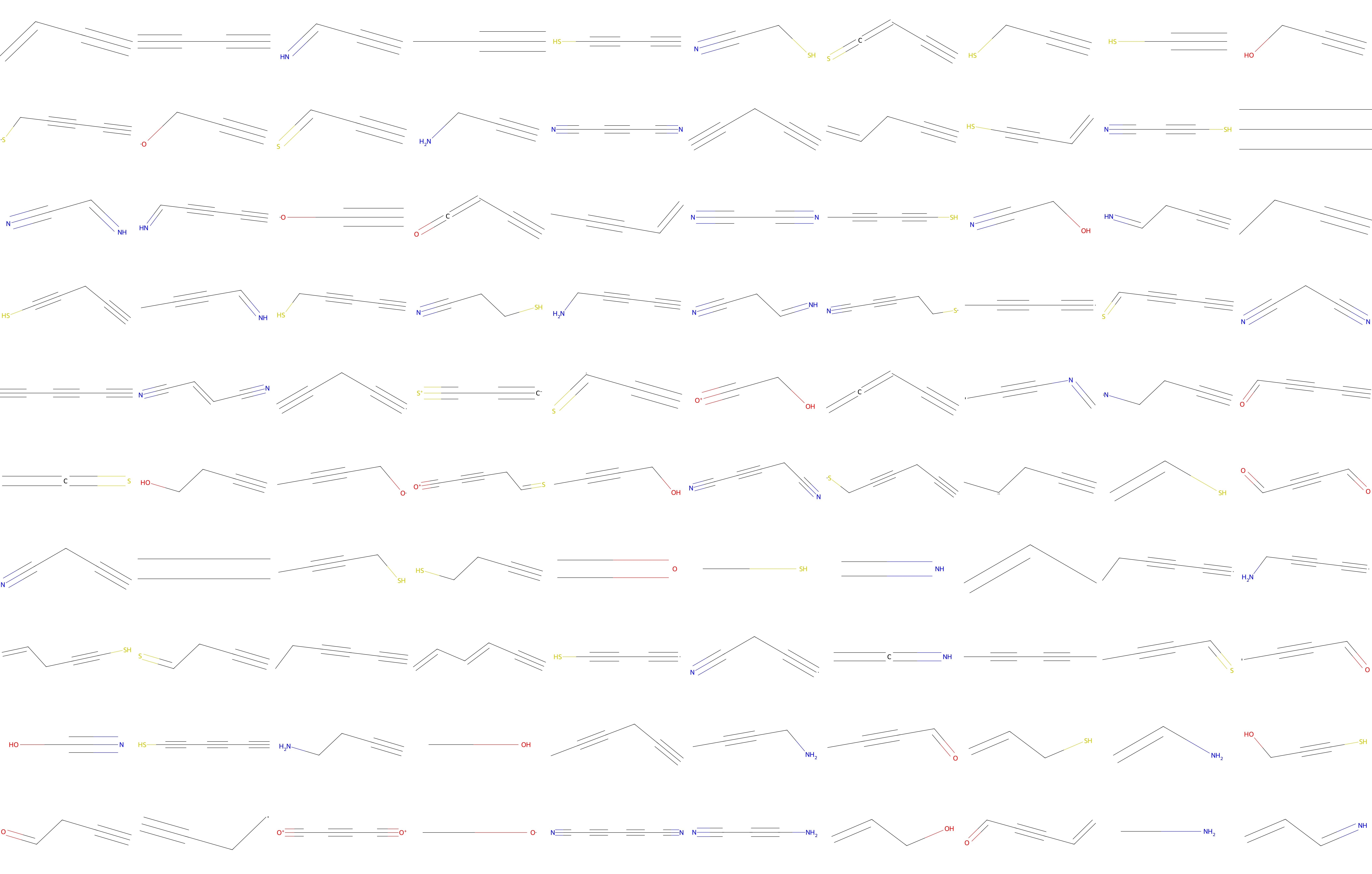} % Adjust width as needed
    \caption{Top 100 ranked candidate molecules for TMC-1 that were generated via the process detailed in Section~\ref{sec:prediction}.}
    \label{fig:predicted_molecules_tmc}
\end{figure*}

\begin{figure*}[htb] % Use figure* for a two-column figure
    \centering
    \includegraphics[width=\textwidth]{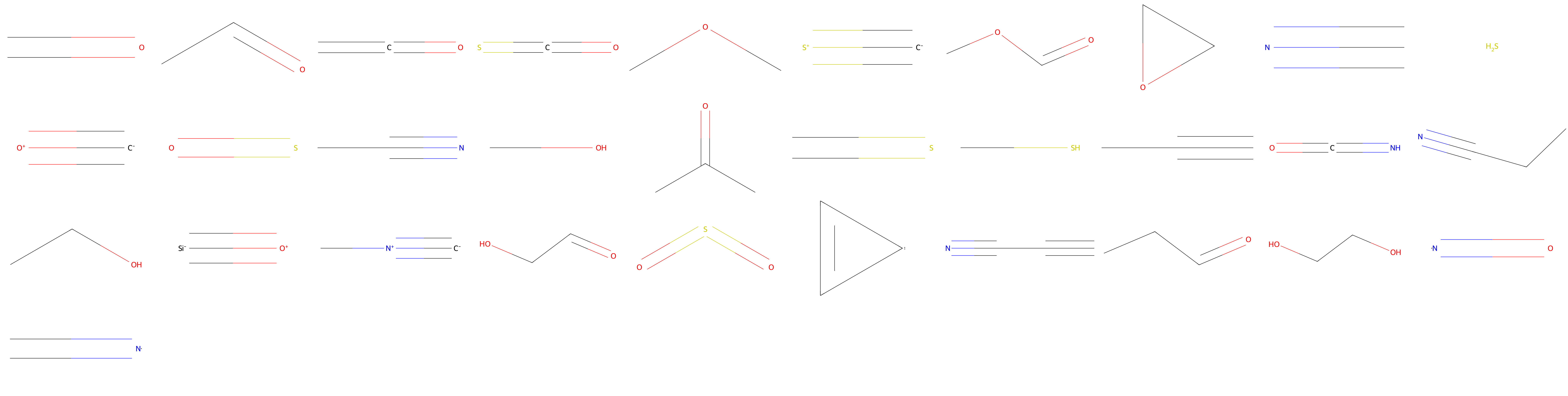} % Adjust width as needed
    \caption{All molecules assigned by the algorithm in the PILS observations of IRAS 16293-2422B, not including isotopically substituted species.}
    \label{fig:assigned_molecules_iras}
\end{figure*}

\begin{figure*}[htb] % Use figure* for a two-column figure
    \centering
    \includegraphics[width=\textwidth]{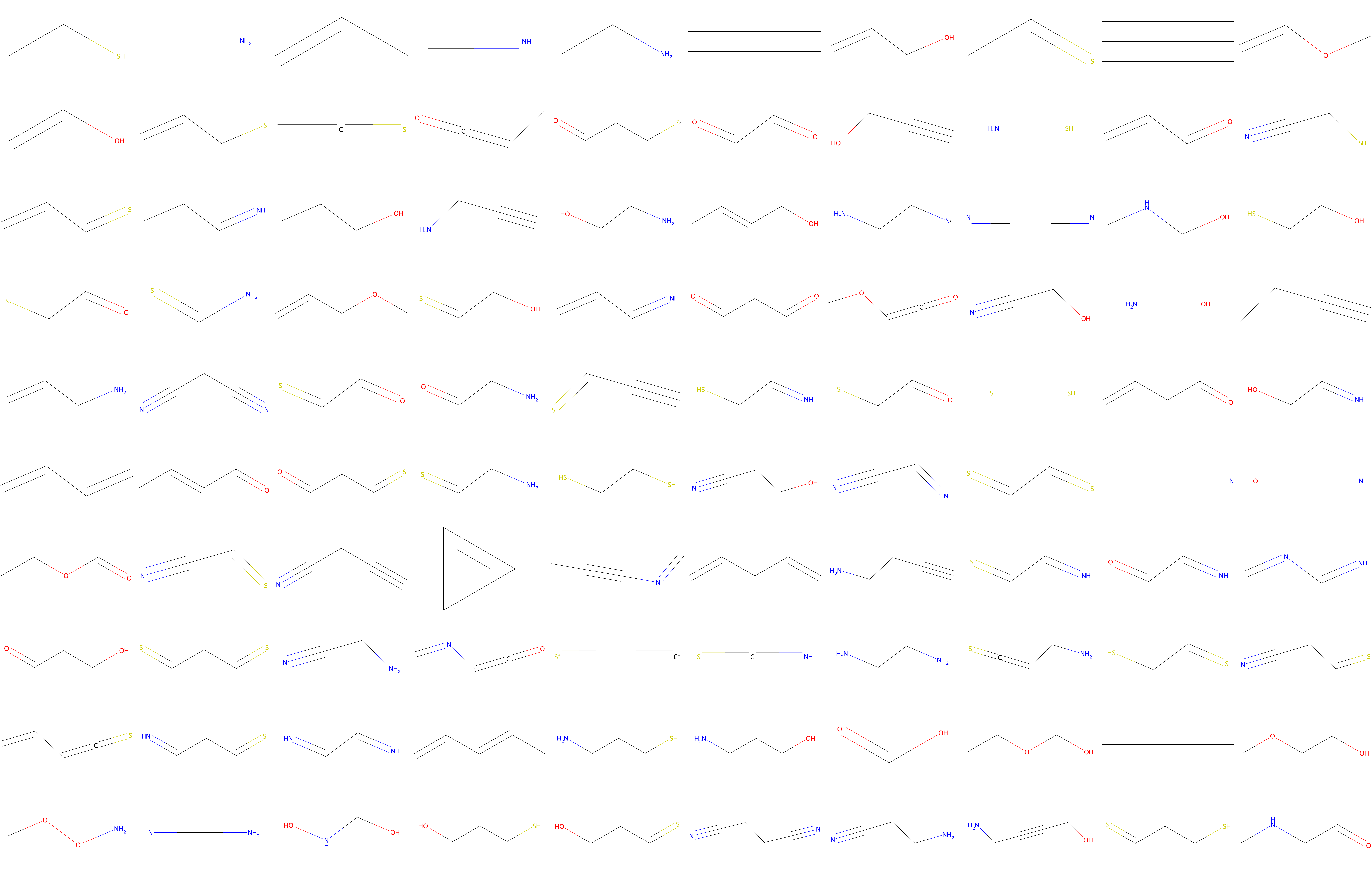} % Adjust width as needed
    \caption{Top 100 ranked candidate molecules for IRAS 16293B that were generated via the process detailed in Section~\ref{sec:prediction}.}
    \label{fig:predicted_molecules_iras}
\end{figure*}

Appendix A contains Figures~\ref{fig:assigned_molecules},\ref{fig:predicted_molecules_tmc}, \ref{fig:assigned_molecules_iras}, and \ref{fig:predicted_molecules_iras} that depict the molecules assigned by the algorithm toward TMC-1 and IRAS 16293B along with those predicted by the generative algorithm.

\section{Appendix B: Details of MLP Trained to Predict Molecular Fragments }
\label{apendix_b}
\begin{figure*}[htb] % Use figure* for a two-column figure
    \centering
    \includegraphics[width=\textwidth]{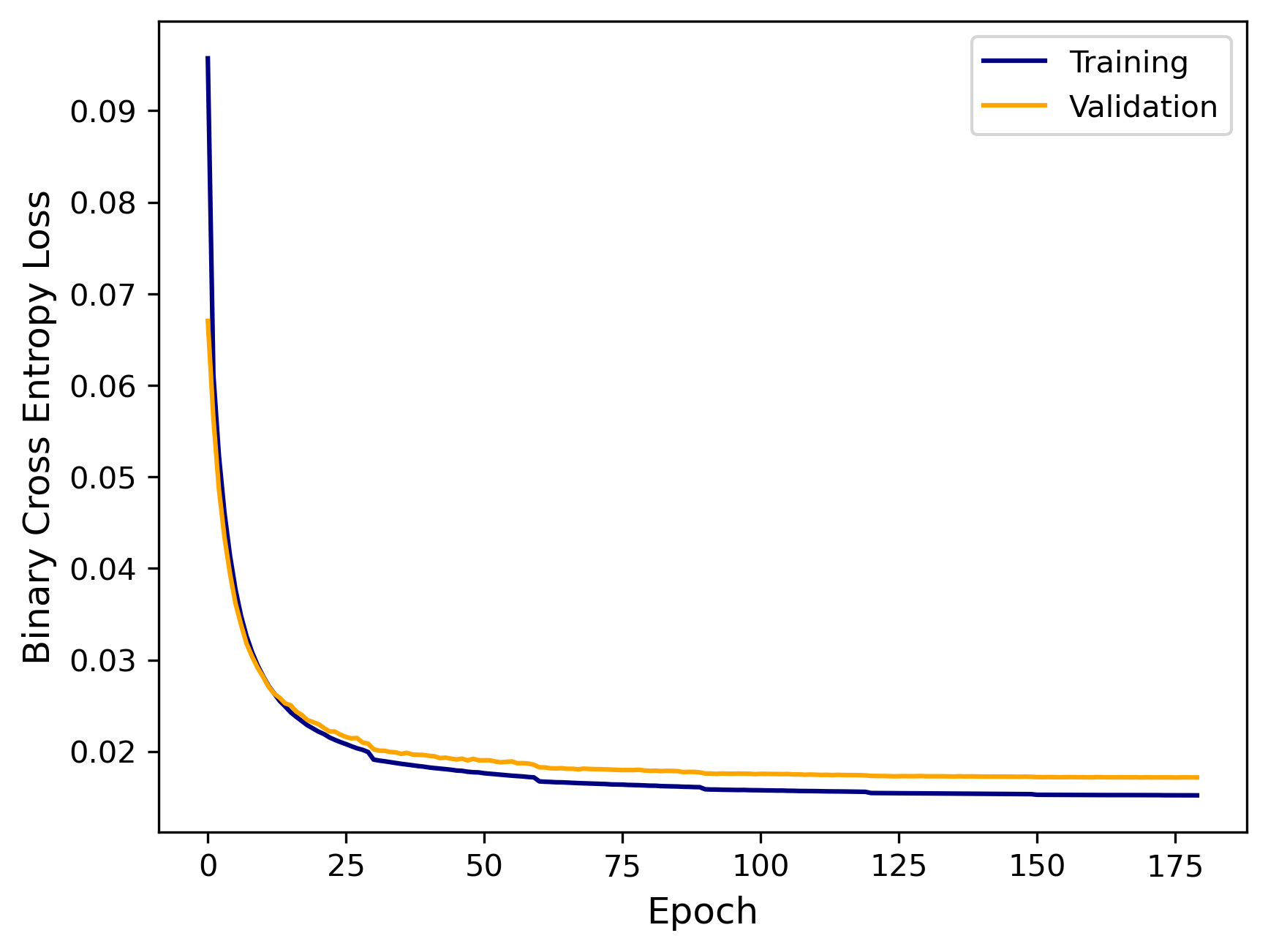} % Adjust width as needed
    \caption{Learning curve of the MLP that was trained to map VICGAE vectors to molecular substructure predictions. Details are provided in Section~\ref{sec:prediction} and Appendix~\ref{apendix_b}}
    \label{fig:learning_curve}
\end{figure*}

The model that was trained to map VICGAE vectors to a 326-dimensional multi-hot-encoded prediction of molecular fragments is a fully connected feedforward neural network with two hidden layers. It takes a 32 dimensional VICGAE vector as its input. Each hidden layer contains 256 nodes. There is a ReLU activation function between each hidden layer, which introduces nonlinearity and allows the network to learn complex relationships. Following the final hidden layer, there is a linear output layer that maps from 256 to 326 nodes, followed by a sigmoid activation function. This activation function results in every output value falling between 0 and 1, representing the probability that the inputted molecule contains the molecular fragment corresponding to that index. The network is trained using binary cross-entropy (BCE) loss, which is common for a classification task where each output represents an independent probability. The data was split 80/20 into training and testing sets. During the training, the learning rate was cut in half every 30 epochs and the training was stopped once the validation loss failed to improve for 10 consecutive epochs. Ultimately, the model converged after 179 epochs. The learning curve is displayed in Figure~\ref{fig:learning_curve}.

To further assess whether the molecular-fragment prediction model provided chemically meaningful results, we evaluated it on a small set of simple molecules that were absent from the training set. For each molecule, we report in Table~\ref{tab:test_fragments} all predicted fragments with model probabilities greater than 20\%, where these probabilities correspond to the output values after the sigmoid activation of the final neural-network layer.

\begin{table*}
\centering
\caption{Predicted fragments for molecules absent from the training set. Only fragments with predicted probabilities greater than 20\% are shown.}
\label{tab:test_fragments}
\begin{tabular}{lll}
\toprule
Molecule & SMILES & Predicted fragments \\
\midrule

Carbonyl sulfide & \texttt{O=C=S} &
\begin{tabular}[c]{@{}l@{}}
R--C(R)=S (99.6\%) \\
R--C(R)=O (27.3\%)
\end{tabular} \\
\midrule

Ethanol & \texttt{CCO} &
\begin{tabular}[c]{@{}l@{}}
R--C(R)(R)--O--R (80.9\%) \\
R--C(R)(R)--C(R)(R)--O--R (63.2\%)
\end{tabular} \\
\midrule

Benzonitrile & \texttt{N\#Cc1ccccc1} &
\begin{tabular}[c]{@{}l@{}}
R--C$\equiv$N (97.6\%) \\
R--C(R)(R)--C$\equiv$N (46.6\%) \\
benzene (21.7\%)
\end{tabular} \\
\midrule

Propyne & \texttt{C\#CC} &
\begin{tabular}[c]{@{}l@{}}
R--C$\equiv$C--R (100\%) \\
R--C$\equiv$C--C(R)(R)(R) (81.1\%)
\end{tabular} \\
\bottomrule
\end{tabular}
\end{table*}

For carbonyl sulfide, the model predicts fragments corresponding to both the C=S and C=O character of the double-bond system. For ethanol, the highest-probability fragments correspond to a basic carbon-alcohol linkage and the full carbon-carbon-oxygen backbone. For benzonitrile, the model predicts fragments corresponding to both the terminal nitrile and aromatic ring components. For propyne, the model predicts alkyne-containing fragments as the dominant structural motifs. Overall, the highest-probability fragments correspond to the expected functional groups or bonding patterns of the held-out molecules, suggesting that the model can identify chemically meaningful features in simple structures absent from the training set.

\section{Appendix C: Fitted Column Densities}
\begin{figure}[h] % Use figure* for a two-column figure
    \centering
    \includegraphics[width=\columnwidth]{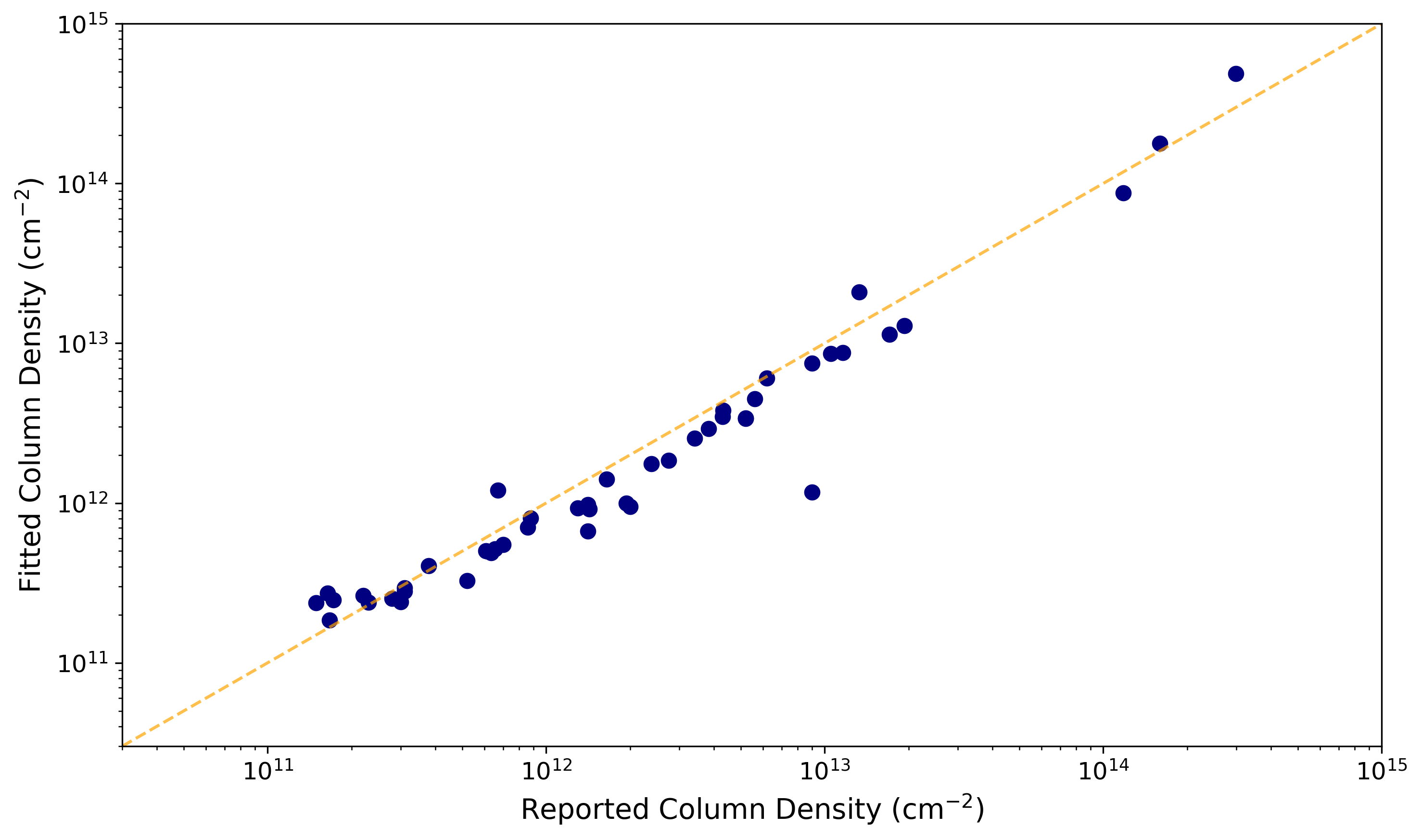} % Adjust width as needed
    \caption{Comparison of fitted column densities from our algorithm with those rigorously determined by \citet{xue_tmc_2025} toward TMC-1. The orange dashed trace represents the $x=y$ line, along which perfectly matched values would lie. As mentioned in the text, the most prominent outlier corresponds to \ce{HCCCHO}, whose sub-thermal excitation is not reproduced by the single-temperature
LTE assumption adopted here.}
    \label{fig:tmc_columns}
\end{figure}

Although the main goal of our code is accurate assignment of molecules in interstellar line surveys, the least-squares fitting procedure yields quick column density estimates for the assigned molecules as well. Recently, \citet{xue_tmc_2025} reported precise column densities for 102 molecular species detected in the GOTHAM survey toward TMC-1 using a Markov-Chain Monte Carlo (MCMC) approach. These reported values by \citet{xue_tmc_2025} are very rigorously and accurately determined and therefore provide a baseline to assess the precision of our values. Figure~\ref{fig:tmc_columns} compares our fitted column densities with those determined by \citet{xue_tmc_2025}. Overall, the agreement is quite strong. A small number of outliers remain, in part because our code assumes a uniform excitation temperature for each molecule. For example, the fitted column density of HCCCHO shows a clear discrepancy because \citet{xue_tmc_2025} and \citet{remijan_propynal_2024} demonstrated that this molecule exhibits sub-thermal excitation in TMC-1 and is best described by an excitation temperature of about 3.1 K. Therefore, assuming a uniform excitation temperature of $\sim$7 K does not reproduce the correct column density. Future work will hopefully allow for several excitation temperatures to be considered.

\begin{figure}[h] % Use figure* for a two-column figure
    \centering
    \includegraphics[width=\columnwidth]{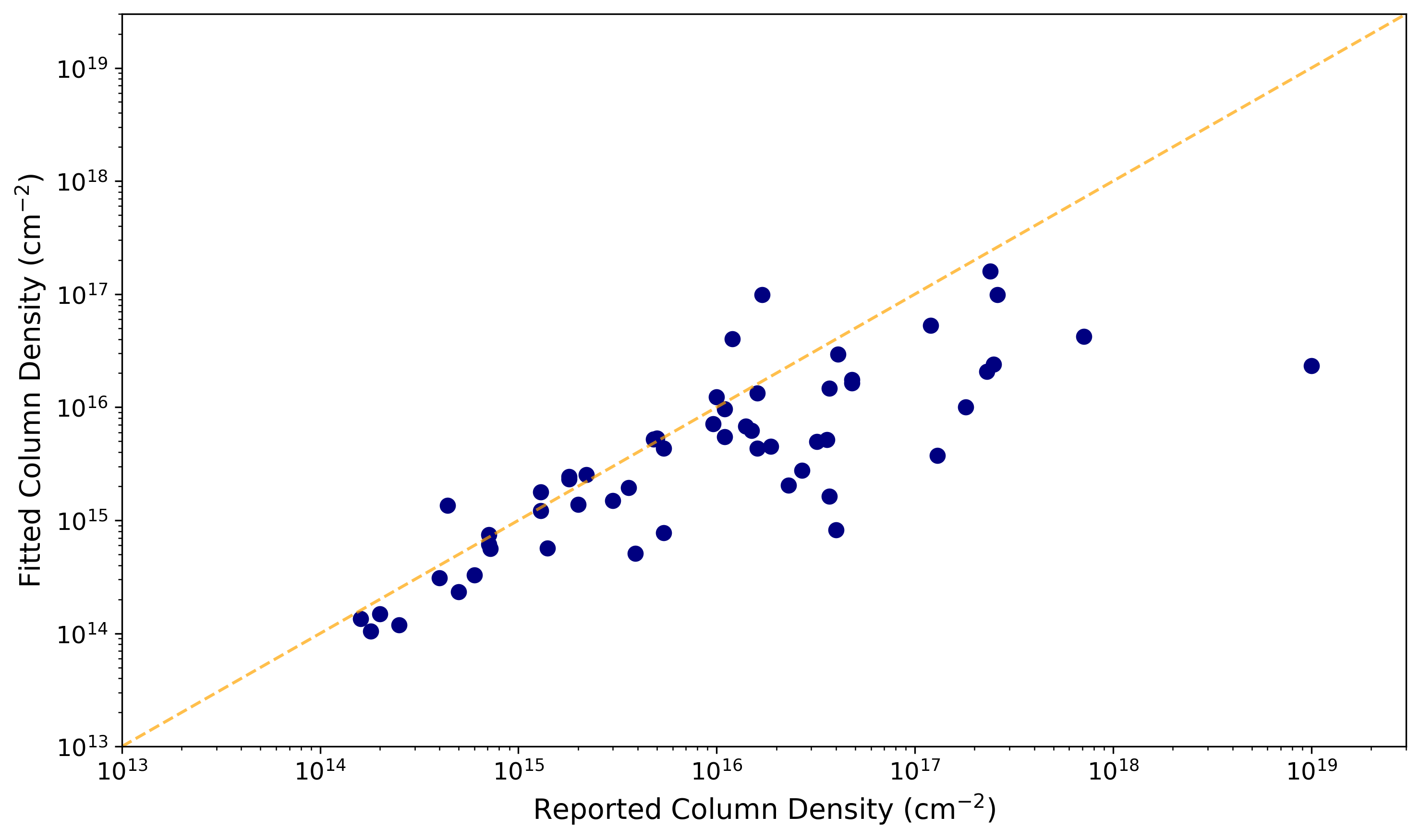} % Adjust width as needed
    \caption{Comparison of fitted column densities from our algorithm with those reported in the literature toward IRAS 16293B. The orange dashed trace represents the $x=y$ line, along which perfectly matched values would lie.}
    \label{fig:iras_columns}
\end{figure}

As shown in Figure~\ref{fig:iras_columns}, the fitted column densities for the molecules in the PILS data show larger deviations from the literature values. This is especially true for some molecules (such as methanol, which corresponds to the rightmost point on Figure~\ref{fig:iras_columns} that strays significantly from the diagonal) that are highly optically thick, making reliable column density estimation difficult. The greater disagreement in this source is somewhat unsurprising since the wider linewidths in the PILS data result in far more blended transitions. Therefore, each fitted column density is typically highly dependent on the fitted values of several other assigned molecules. Additionally, the excitation temperatures used in previous studies of the PILS data vary considerably among molecules \citep{jorgensen_alma-pils_2018}, whereas in our analysis, each species is assumed to be characterized by a single excitation temperature. Finally, the partition functions used to simulate each molecule’s spectrum may omit the vibrational or conformational corrections applied in some literature column-density determinations, which could introduce additional discrepancies.

Overall, while our least-squares fitting may provide a rough initial approximation, the resulting column densities would benefit from refinement using more rigorous fitting methods. 

\end{document}